# Nonlinear pressure dependence of $T_N$ in almost multiferroic EuTiO$_3$

Z. Guguchia[1], K. Caslin[2], R. Kremer[2], H. Keller[1], A. Shengelaya[3], A. Maisuradze[1], J. L. Bettis Jr.[4], J. Köhler[2], A. Bussmann-Holder[2], and M.-H. Whangbo[4]

[1] Physik-Institut der Universität Zürich, Winterthurerstr. 190, CH-8057 Zürich, Switzerland
[2] Max-Planck-Institute for Solid State Research, Heisenbergstr.1, D-70569 Stuttgart, Germany
[3] Department of Physics, Tbilisi State University, Chavchavadze 3, GE-0128 Tbilisi, Georgia
[4] Department of Chemistry, North Carolina State University, Raleigh, NC 27695-8204, USA

**Abstract**
The antiferromagnetic (AFM) phase transition temperature $T_N$ of EuTiO$_3$ has been studied as a function of pressure $p$. The data reveal a nonlinear dependence of $T_N$ on $p$ with $T_N$ increasing with increasing pressure. The exchange interactions exhibit an analogous dependence on $p$ as $T_N$ (if the absolute value of the nearest neighbor interaction is considered) and there is evidence that the AFM transition is robust with increasing pressure. The corresponding Weiss temperature $\Theta_W$ remains anomalous since it always exhibits positive values. The data are analyzed within the Bloch power law model and provide excellent agreement with experiment.

Pacs-Index: 75.50.Ee, 75.30.Kz, 61.50.Ks

## 1. Introduction

EuTiO$_3$ (ETO) has recently attracted novel interests due to its possible multiferroic behavior and anomalously strong spin – lattice coupling [1 – 3]. At $T_N$ = 5.5 K the compound undergoes a paramagnetic to antiferromagnetic (AFM) phase transition [4]. Over a large temperature regime a transverse optic long wave length mode softens reminiscent of a ferroelectric phase transition [2, 3]. Complete softening is inhibited by quantum fluctuations [5, 6] as is also observed for SrTiO$_3$ (STO) and other perovskite oxides. Upon entering the AFM state this mode experiences an unexpected hardening in ETO demonstrating strong spin – lattice coupling [1 – 3]. The obvious analogy between ETO and STO has been extended recently by predicting and verifying experimentally that ETO also exhibits an oxygen octahedral rotational instability [7], however, at much higher temperature $T_S$ ($T_S$ = 282K) than in STO ($T_S$ = 105K). This large difference in transition temperatures was a motivation for further studies not only on ETO but also of the mixed crystals ETO-STO [8, 9]. From these investigations the phase diagram for this series has been established with $T_S$ depending nonlinearly on STO content [9]. In a pure system ETO novel dynamics have been observed via muon spin rotation (μSR), namely, that at temperature $T_N < T_S < T$ a finite paramagnetic μSR signal is present which must stem from spin correlated regions with finite spatial extent [10]. This result is further evidence further that an unusual spin – lattice coupling exists in ETO which is established at high temperatures. This interpretation of the data has been verified by demonstrating that $T_S$ is strongly dependent on the magnetic field [11], a feature so far unknown in nominally paramagnetic insulators. In order to characterize this interesting system further, the low temperature Néel state is investigated by applying pressure and measuring the $p$ dependence of $T_N$. Here we emphasize that our interest is in the bulk magnetic and structural properties of ETO only which are distinctly different from thin films or substrates. These have been the focus of numerous papers with strain and stress engineering of material properties and are beyond the current investigation. It is also important to mention in this context that in quasi-two-dimensional (2D) materials their physics is very different from that observed for their three-dimensional (3D) analogues, which makes it impossible to discuss phenomena observed in these 2D compounds in relation to the bulk materials. The low temperature phase has already been addressed experimentally via different approaches [12, 13], namely by investigating its magnetic and electric field dependences. From these studies it is concluded that the ground state of ETO is a multidomain state with the possibility of developing ferroelectricity if symmetry breaking can be achieved.

## 2. Methodology

Here we apply pressure to ETO to test the stability range of the AFM state and explore the possibility to achieve a ferromagnetic (FM) state. This is motivated by the fact that in semiconducting cubic Eu chalcogenides the systems change their magnetic states from FM to AFM with increasing ionic radius [14]. This observation corresponds to an inverse pressure effect which offers the possibility that ETO can be transformed from AFM to FM with increasing pressure. On the other hand pressure experiments on various AFM perovskites and spinels [15, 16] and Ce containing compounds [17 – 19] have evidenced that $T_N$ is stabilized and increases with increasing pressure. Calculations for ETO within a Landau-Ginzburg free energy expansion and *ab initio* computations support the possibility to achieve FM order in ETO [20, 21] which is in accordance with first principles calculations [8]. These have demonstrated that the AFM and FM ground states have almost the same energy with an energy gain of a few meV in favor of AFM ordering. As such it appears timely to establish the pressure dependence of the Néel state for ETO. The data are analyzed within the Bloch power law model [22] which has already been employed for numerous other FM and AFM systems and proven to be particularly useful [14]. A comparison of these model results with those derived from Monte Carlo studies for the Eu chalcogenides has demonstrated its outstanding qualification especially when considering pressure effects on AFM and FM states [14].

The ETO powder samples used have been prepared as described in [7]. Measurements of the temperature dependence of the magnetic moment $m$ for the sample $EuTiO_3$ were performed with a commercial SQUID magnetometer (*Quantum Design* MPMS-XL). Investigations were carried out at ambient as well as under applied pressures up to $p = 57$ kbar using a diamond anvil cell (DAC) [23] filled with Daphne oil which served as a pressure-transmitting medium. The pressure at low temperatures was determined by the pressure dependence of the superconducting transition temperature of Pb.

## 3. Results

The temperature dependence of the magnetic susceptibility $\chi$ for $EuTiO_3$ recorded at ambient pressure is shown in figure 1(a) with the background signal of the empty pressure cell being subtracted. $T_N$ is clearly visible as a distinct peak in the susceptibility data marked by the arrow and $T_N = 5.5K$ Below $T_N$ a slight increase in χ takes place which might be caused by insufficient background subtraction and small amounts of paramagnetic impurities. From the inverse susceptibility (inset to figure 1(a)) the Weiss temperature $\Theta_W = 3.4$ K is obtained in agreement with previous data [4]. The field dependence of the magnetization confirms a gradual change from AFM to FM with increasing magnetic field and saturation is achieved for fields larger than 2T with a saturation magnetization of 6.73 $\mu_B$ which is very close to the spin magnetic moment of $Eu^{2+}$ (7 $\mu_B$). Both of these results are in agreement with data for $EuZrO_3$ which becomes AFM at a slightly lower temperature $T_N = 4.1K$ [24].

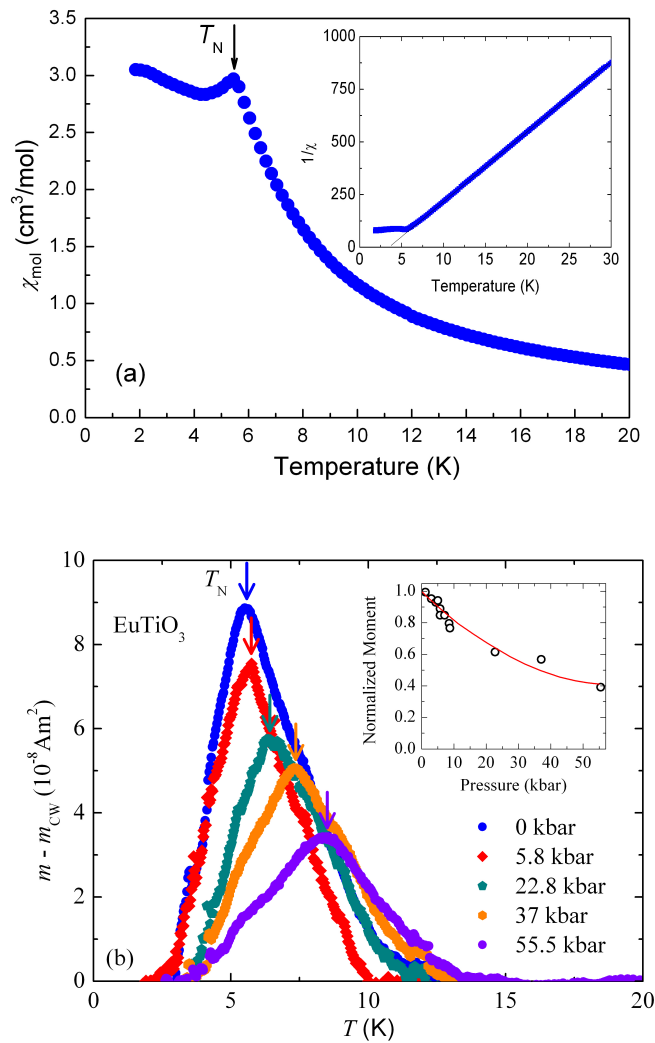

**Figure 1.** (a) Temperature dependence of the magnetic susceptibility χ for ETO at ambient pressure. In the inset the temperature dependence of the inverse susceptibility 1/χ is shown from which the Weiss temperature $\Theta_W = 3.4K$ has been derived. (b) Temperature dependence of the magnetic moment difference $m - m_{CW}$ at ambient and selected applied pressures. The inset shows the normalized peak maximum as a function of pressure. Arrows mark the magnetic ordering temperatures of the Eu moments.

In order to highlight the corresponding peak in the pressure dependent magnetization data more clearly, we subtract form the magnetic moment $m(T)$ the Curie-Weiss type temperature dependence $m_{CW} = C/(T-\Theta_W)$ which confirms a well defined peak at $T_N$ = 5.5 K in $m(T) - m_{CW}(T)$. In figure 1(b) this difference is shown at ambient and selected applied pressures.

The pressure dependence of $T_N$ is shown in figure 2. As is obvious from figure 2 $T_N$ increases nonlinearly with pressure for $p > 10$ kbar. Below this pressure a linear increase in $T_N$ with pressure appears (inset to figure 2). Such behavior has been observed in various other perovskites and also in linear chain antiferromagnets [15 – 19]. The interpretation of those data was based on the fact that the 4f-4f overlap increases with increasing pressure and stabilizes the AFM order, whereas the superexchange via the bridging oxygen ions is considered to be less effective. Figure 1(b) illustrates that with increasing pressure the peak height at $T_N$ diminishes nonlinearly (inset to figure 1(b)) and peak itself broadens. A similar observation has been made in $CeFe_2$ alloys and been interpreted as enhancement of AFM correlations [19]. In the case of ETO this corresponds to an increasing Eu 4f hybridization effect stabilizing the AFM nearest neighbor exchange.

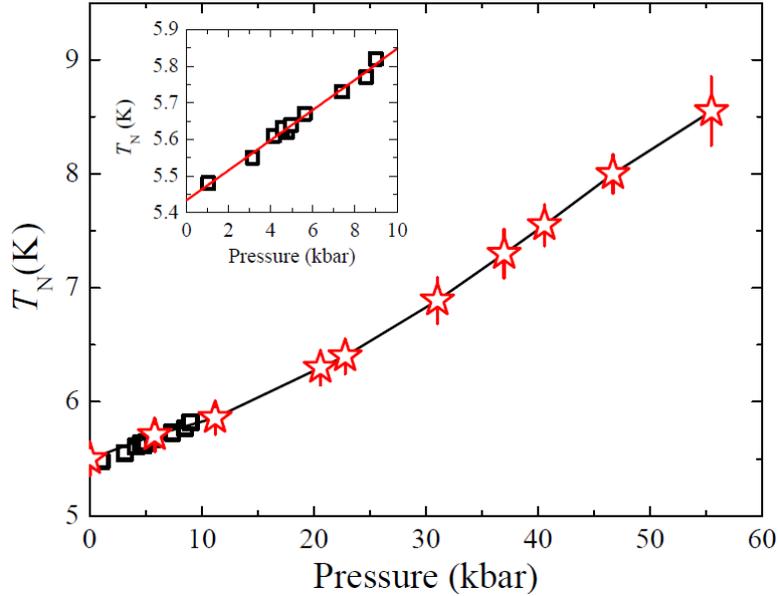

**Figure 2.** Pressure dependence of $T_N$. The inset shows the low pressure dependence of $T_N$ as obtained by SQUID magnetization measurements using a Cu-Be clamp cell ($p < 12$ kbar), also shown as open squares in the main figure. The line is derived theoretically as discussed in the text.

## 4. Discussion

Interestingly, in most cases it has been possible to explain the data in terms of the Bloch power law, where the exchange constants adopt an interatomic distance dependence in terms of the magnetic Grüneisen power laws. In particular, for the Eu chalcogenides a convincing agreement with pressure data could be achieved and the validity of this law confirmed by Monte Carlo studies [14]. This has been taken as motivation to analyze our data within the same framework. Since the bulk modulus of ETO is unknown, we have first established the pressure dependence of $T_N$ in ETO by using the Heisenberg Hamiltonian to calculate $T_N$:

$$T_N = \frac{2S(S+1)}{3k_B}(-6J_1 + 12J_2) \qquad (1)$$

where $J_1$ is the nearest neighbor direct Eu – Eu AFM exchange interaction and $J_2$ the second nearest neighbor indirect ferromagnetic exchange interaction, and $S = 7/2$ being the Eu spin. By adopting the following power laws for the exchange interactions:

$$\tilde{J}_1(p) = J_1(a/a_0)^{-n_1} \text{ and } \tilde{J}_2(p) = J_2(a/a_0)^{-n_2} \qquad (2)$$

with $a$ being the pressure dependent lattice constant and $a_0$=3.904Å the pseudocubic lattice constant at ambient pressure, with $n_1$=20.9 and $n_2$=10.8 being consistent with the Grüneisen exponents, and $J_1/k_B$=-0.0167K and $J_2/k_B$=0.0355K, the pressure dependence of $T_N$ is correctly reproduced (black line in figure 2). Note, that similar values for the exponents of $J_1$ and $J_2$ have been derived in [25]. From this methodology the pressure dependence of the lattice constant can be derived which is shown in figure 3. Since the data have been taken at low temperatures, ETO is in the tetragonal phase. However, with the tetragonal distortion being very small [12, 13, 26], the pseudocubic lattice constant is used and plotted as a function of pressure. This shrinks linearly with pressure. Its $p$ dependence is comparable to the one of STO [27, 28], however it is slightly steeper. From this dependence the spontaneous strain $e_1 = e_2 = (a(p) - a_0)a_0$ is calculated (lower inset in figure 3). As compared to STO, ETO develops a larger strain with increasing $p$ and a similar evolution of the relative volume change (upper inset in figure 3). But the general behavior for all three quantities is qualitatively the same as in STO [28].

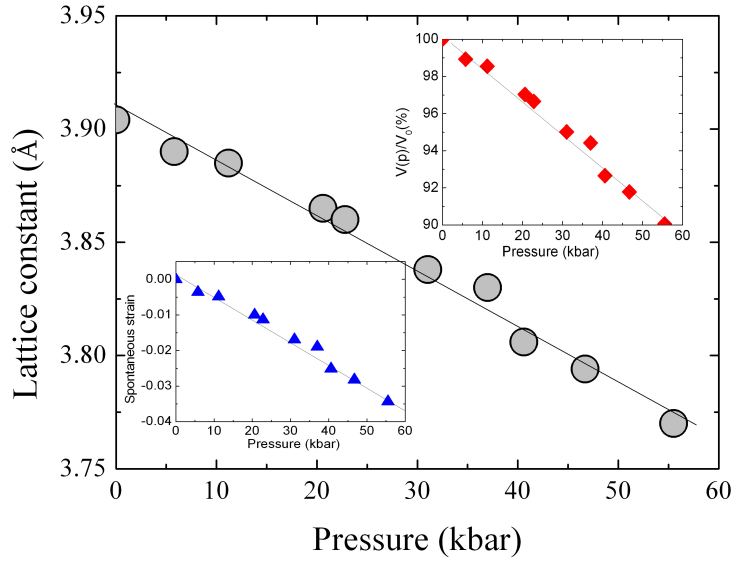

**Figure 3.** Pressure dependence of the pseudo-cubic lattice constant for ETO. The upper inset shows the normalized volume as a function of pressure, whereas the lower inset displays the pressure dependence of the spontaneous strain.

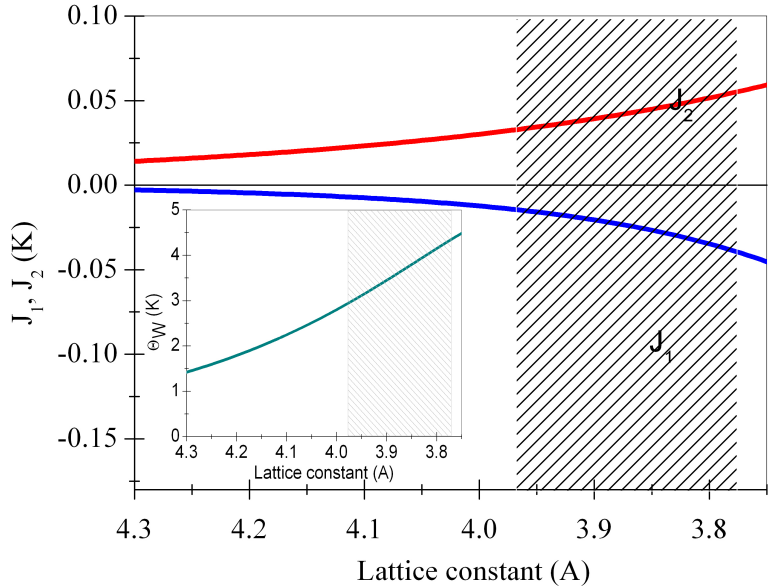

**Figure 4.** Dependence of the nearest ($J_1$, blue curve) and next nearest neighbor ($J_2$, red curve) exchange constants on the pseudo-cubic lattice constant for ETO. The shaded area refers to the experimentally accessible region. The inset shows the Weiss temperature $\Theta_W$ as a function of the pseudo-cubic lattice constant. The experimentally accessible region for the lattice constant compression and dilatation has been largely exaggerated in order to demonstrate the nonlinear dependence of $\Theta_W$ on $a$. The shaded area refers to the experimentally accessible region.

The nearest and next nearest neighbor exchange constants as derived from equation (2) are shown in figure 4 as a function of the pressure dependent lattice constant. While $J_2$ increases steadily with decreasing lattice constant (increasing pressure) $J_1$ decreases within the same range, supporting the AFM order. This trend for $J_1$ is supported by LDA+U calculations where a small energy gain in favor of G-type AFM order is achieved as compared to FM order [29]. If the system could, however, be artificially tensile stressed and the lattice constant enlarged by 14% a sign change of $J_1$ takes place enabling a transition to a FM state. Such large tensile stresses are experimentally not achievable and correspondingly a transition from AFM to FM can be excluded. The experimentally accessible range of $a$ is highlighted in Figure 4 by the shaded area. Interestingly, however, recent amorphization of the ETO samples has been shown to result in FM order at about the same temperature as AFM order is established in the crystalline sample [30, 31]. This has been attributed to a volume expansion together with the 5d magnetic polarization of the $Eu^{2+}$ with the former being consistent with the trends predicted here. Since these data have been obtained on thin films, a direct comparison to our analysis is not possible. It is important to note in addition, that also for the compound $Eu_{0.5}Ba_{0.5}TiO_3$ the lattice constant is enlarged by 1.13% as compared to ETO [32]. Here magnetic susceptibility data suggest a possible FM behavior with the Curie temperature being below 4K. Since, as outlined above, this tensile strain is insufficient to induce FM order, the dilution of Eu moments due to Ba substitution could be the cause of such a transition. On the other hand, in more recent experiments on the same composition samples [33], lower temperatures than used in [32] could be attained and a transition to AFM order seen at $T_N = 1.9$ K, which is rather consistent with our conclusions from figure 3. In tensile and compressive biaxial strain engineered films of ETO a transition to FM order has been observed [34] at 1 % strain only. In this case, which cannot be compared to our bulk samples, as outlined above, it is very likely that the stronger reduction of nearest neighbor spin-spin interactions as compared to the second nearest neighbors (in strictly 2D $z_1 = 5$, $z_2 = 8$) is the reason for the appearance of a FM state with a much reduced strain that suggested from our analysis.

From the pressure dependence of the exchange constants the pressure dependent Weiss temperature $\Theta_W$ is derived. As has been demonstrated before, $\Theta_W$ is anomalous not only in pure ETO [25] but also in the STO-ETO mixed crystals since it is positive, as a consequence of the vicinity of ETO to FM order [8, 9]. By plotting $\Theta_W$ as a function of the lattice constant (exaggerating the possible scale of *a*) it is seen that AFM order is supported by pressure (inset to figure 4). Tensile strain which would cause an increase in the lattice constant, on the other hand also leads to a decrease in the Weiss temperature; here, however, a sign change does not take place. Since such methodologies can only be performed on thin films, we cannot compare this with the bulk material and its hydrostatic pressure dependence. Overall, the dependence of $\Theta_W$ on *a* is anomalous since nonlinear behavior is observed, caused by the competing exchange interactions and their different power law dependences.

## 5. Conclusion

In conclusion, we have investigated the pressure dependence of the low temperature paramagnetic-antiferromagnetic phase transition for ETO. For low pressures a linear increase of $T_N$ with *p* is observed which adopts a nonlinear dependence with higher *p*. The data have been analyzed within the Bloch power law model from which the pressure dependence of the pseudo-cubic lattice constant has been derived. Both exchange constants, namely nearest and next nearest neighbor, decrease / increase with increasing *p*, respectively, thus not allowing for the appearance of a FM state. On the other hand exceedingly large tensile strain offers the possibility for a FM state, however, with values orders of magnitude beyond those experimentally achievable. The Weiss temperature remains positive in all cases demonstrating the unusual nature of the AFM state. Remarkable is ETO's highly nonlinear evolvement of $T_N$ with *p* which has no comparable analogs.

**Acknowledgments**

This work is partly supported by the Swiss National Science Foundation, the NCCR Program MaNEP, and a SCOPES Grant No. IZ73Z0_128242.